\documentclass[twocolumn]{aastex631}

\usepackage{graphicx}
\usepackage{amsmath}
\usepackage{subfigure}
\usepackage{multirow}
\usepackage{threeparttable}
\usepackage{array}
\usepackage{natbib}
\usepackage{bm}
\usepackage{soul}

\newcommand{\tgc}{\theta_{\rm GC}}
\newcommand{\xmm}{{\it XMM-Newton}}

\defcitealias{Pan:2024aa}{Paper I}

\begin{document}
\title{The \textbf{\textit{XMM-Newton}} Line Emission Analysis Program (X-LEAP) II: The Multi-scale Temperature Structures in the Milky Way Hot Gas}

\correspondingauthor{Zhijie Qu}
\email{quzhijie@uchicago.edu}

\author[0000-0002-2941-646X]{Zhijie Qu}
\affiliation{
Department of Astronomy $\&$ Astrophysics, The University of Chicago, Chicago, IL 60637, USA}

\author[0000-0002-9943-1155]{Zeyang Pan}
\affiliation{Key Laboratory of Optical Astronomy, National Astronomical Observatories, Chinese Academy of Sciences, 20A Datun Road, Beijing, 100101, People's Republic of China}
\affiliation{School of Astronomy and Space Science, University of Chinese Academy of Sciences, Beijing, 100049, People's Republic of China}

\author[0000-0001-6276-9526]{Joel N. Bregman}
\affiliation{Department of Astronomy, University of Michigan, Ann Arbor, MI 48109, USA}

\author{Jifeng Liu}
\affiliation{Key Laboratory of Optical Astronomy, National Astronomical Observatories, Chinese Academy of Sciences, 20A Datun Road, Beijing, 100101, People's Republic of China}
\affiliation{School of Astronomy and Space Science, University of Chinese Academy of Sciences, Beijing, 100049, People's Republic of China}
\affiliation{National Astronomical Observatories,
Chinese Academy of Sciences, 20A Datun Road, Chaoyang District, Beijing, China}
\affiliation{University of Chinese Academy of Sciences, Yuquan Road, Shijingshan District, Beijing, China}



\begin{abstract}
This paper presents the multi-scale temperature structures in the Milky Way (MW) hot gas, as part of the {\it XMM-Newton} Line Emission Analysis Program (X-LEAP), surveying the \ion{O}{7}, \ion{O}{8}, and Fe-L band emission features in the \xmm~ archive.
In particular, we define two temperature tracers, $I_{\rm OVIII}/I_{\rm OVII}$ (O87) and $I_{\rm FeL}/(I_{\rm OVII}+I_{\rm OVIII})$ (FeO).
These two ratios cannot be explained simultaneously using single-temperature collisional ionization models, which indicates the need for multi-temperature structures in hot gas.
In addition, we show three large-scale features in the hot gas: the eROSITA bubbles around the Galactic center (GC); the disk; and the halo.
In the eROSITA bubbles, the observed line ratios can be explained by a log-normal temperature distribution with a median of $\log T/{\rm K} \approx 6.4$ and a scatter of $\sigma_T \approx 0.2$ dex. 
Beyond the bubbles, the line ratio dependence on the Galactic latitude suggests higher temperatures around the midplane of the MW disk.
The scale height of the temperature variation is estimated to be $\approx$2 kpc assuming an average distance of $5$ kpc for the hot gas.
The halo component is characterized by the dependence on the distance to the GC, showing a temperature decline from $\log\,T/{\rm K}\,\approx\, 6.3$ to $5.8$.
Furthermore, we extract the auto-correlation and cross-correlation functions to investigate the small-scale structures.
O87 and FeO ratios show a consistent auto-correlation scale of $\approx$$ 5^\circ$ (i.e., $\approx$$ 400$ pc at 5 kpc), which is consistent with expected physical sizes of X-ray bubbles associated with star-forming regions or supernova remnants.
Finally, we examine the cross-correlation between the hot and UV-detected warm gas, and show an intriguing anti-correlation. 
\end{abstract}



\section{Introduction}
\label{sec:intro}

The massive gaseous halo surrounding a galaxy is formed during galaxy formation and evolution, which is also known as the circumgalactic medium (CGM; see recent reviews \citealt{Donahue:2022aa} and \citealt{Faucher-Giguere:2023aa}).
Among the multiphase medium in the gaseous halo, the hot gas is a dominant contributor to the total baryon mass $\gtrsim$$ L^*$ galaxies. \citep[e.g.,][]{Bregman:2018aa}.
The hot halo plays a fundamental role in modulating galaxy growth by preventing direct accretion from the intergalactic medium (IGM), providing continuous accretion through radiative cooling, and gathering the feedback materials ejected from the galaxy \citep[e.g.,][]{Cen:2006aa, Keres:2009aa}.

Observations in the past decade revealed the Milky Way (MW) is surrounded by the diffuse hot gas \citep[e.g.,][]{Henley:2012aa, Gupta:2012aa, Miller:2013aa, Miller:2015aa, Nakashima:2018aa, Kaaret:2020aa, Ponti:2023aa, Zheng:2024aa}.
Around the Galactic center (GC), two X-ray bubbles are discovered in the eROSITA all-sky survey \citep{Predehl:2020aa}.
In addition, the density distribution of the hot gas disk and halo is constrained by considering the X-ray emission or absorption dependence on the Galactic longitudes ($l$) and latitudes ($b$).
However, the decomposition of the disk and halo in observations is still uncertain.
Previous studies obtain different density distributions using data obtained from different instruments (e.g., \xmm, {\it Suzaku}, or {\it HaloSat}; e.g., \citealt{Li:2017aa, Nakashima:2018aa, Kaaret:2020aa}).

The degeneracy of the density and temperature distribution in X-ray emission or absorption makes it harder to accurately determine the density distribution.
For example, assuming a constant temperature, modeling found the hot halo density distribution can be approximated by a power law with a slope of $-1.5$, leading to a total mass of $\approx 4-5 \times 10^{10}~ \rm M_\odot$ within 250 kpc \citep[e.g.,][]{Li:2017aa}.
Assuming hydrostatic equilibrium (i.e., declining temperature to the outskirts), the density distribution is flatter in the outer halo, leading to a more massive hot halo of $\gtrsim$$ 10^{11}~ \rm M_\odot$ \citep[e.g.,][]{Faerman:2017aa}.
This discrepancy is fundamental in resolving the missing baryons problem in the MW, where $\approx$$ 50\%$ of baryons ($\approx$$ 10\times 10^{10}\rm ~M_\odot$) are expected beyond the MW disk \citep[e.g.,][]{McGaugh:2010aa}.

The majority of the MW hot gas has a temperature consistent with the virial temperature of $1-2\times 10^6$ K expected for an $L^*$ galaxy.
Previous studies found that the hot gas temperature is roughly constant over the entire sky \citep[e.g.,][]{Henley:2013aa, Kaaret:2020aa}.
One exception is around the GC (i.e., the Fermi bubbles, FBs; \citealt{Su:2010aa}), where the FBs and shocked shells have a high temperature of $4-5\times 10^6$ K \citep{Miller:2016aa}.
More recently, the hotter gas surrounding the FBs has been found to form larger X-ray bubbles, the eROSITA bubbles \citep{Predehl:2020aa}.
Besides the spatial variation of the temperature, a super-virial hot phase with a temperature of $\approx$$ 5-10 \times 10^6$ K has been reported by recent studies (e.g., \citealt{Das:2019aa, Das:2019ab, Bluem:2022aa, Ponti:2023aa}). 
This extremely hot phase is supported by the enhanced emission feature at $\log T/{\rm K}\approx 6.7-7.0$ or absorption features of highly ionized species (e.g., \ion{Ne}{9} and \ion{Ne}{10}; \citealt{Das:2019ab}).
These existing studies reveal the complexity of the temperature structures in the MW hot gas.

Aiming to understand the MW hot gas, we conduct the {\it XMM-Newton} Line Emission Analysis Program program (X-LEAP), a deep survey of prominent X-ray emission features using the \xmm\, archive \citep{xmm}.
The full introduction to the entire program is presented in \citet[][hereafter \citetalias{Pan:2024aa}]{Pan:2024aa}.
In this study, we report the empirical constraints on the temperature variations in the MW hot CGM, including both the global profiles and small-scale variations.
In Section \ref{sec:data_reduction}, we briefly introduce the sample, data reduction, and emission feature extraction.
Section \ref{sec:results} presents the major results including the temperature constraints of the eROSITA bubbles (\S \ref{sec:bubbles}), the disk component (\S \ref{sec:disk}), and the halo (\S \ref{sec:halo}).
The small-scale structures are investigated using the auto-correlation and cross-correlation functions (\S \ref{sec:acf}).
The implications of this work are discussed in Section \ref{sec:dis}.
In the end, the key findings are summarized in Section \ref{sec:summary}.

\section{sample and data reduction}
\label{sec:data_reduction}

In \citetalias{Pan:2024aa}, we reduce all observations in the \xmm~ archive over the past two decades ahead of 2022, yielding a final sample of 5418 observations. 
This sample is optimized to study both the temporal and spatial variations of the diffuse soft X-ray emission.
In this work, we directly use the data reported in \citetalias{Pan:2024aa}, and provide a brief overview of the sample selection and data reduction in this section.

\subsection{The \xmm\, archival sample}
In the past two decades, the \xmm~ archive accumulated more than 543 Ms of observations covering $4.49\%$ of the entire sky ($1768 ~{\rm deg}^2$).
Compared to all-sky surveys (e.g., ROSAT and eROSITA), \xmm~ observations are deeper with typical exposures of $\approx$$ 20$ ks for each pointing, which is one order of magnitude longer than the designed depth of eROSITA of $2.5$ ks \citep{eROSITA}.
Because \xmm~ and eROSITA exhibit similar effective areas at $\approx 0.5-1$ keV \citep{mos, eROSITA}, one order of magnitude longer exposures in the \xmm~ archive leads to a limiting flux $\approx 3$ times lower than eROSITA in the overlapped sky regions.
Therefore, the \xmm~ archive is more powerful in decomposing different components in the soft X-ray emission (e.g., the Galactic hot gas emission and cosmic X-ray background).
In the X-LEAP program, we focus on three spectral features, the \ion{O}{7} K$\alpha$ triplet at $\approx$$ 560$ eV, \ion{O}{8} K$\alpha$ at $\approx$$ 654$ eV, and Fe L-shell band (Fe-L) at $\approx$$ 800$ eV.
Each spectral feature has multiple transitions with similar ionization states and emission energies, which are defined in \citetalias{Pan:2024aa}.
These three spectral features are sensitive to gas at different temperatures \citep{Foster:2012aa, Del-Zanna:2015aa}.
In the collisional ionization equilibrium (CIE) model, \ion{O}{7}, \ion{O}{8}, and Fe-L band emission peak at $2\times 10^6$ K, $3\times 10^6$ K, and $7\times 10^6$ K, respectively.

In the \xmm~ archive, there are 15035 individual observations obtained from 2000 to 2022.
In the final sample, we selected 5418 observations after omitting those with short exposure times or strong contaminations.
All selected observations have effective exposure times of $>$$ 5$ ks  (i.e., the good time intervals obtained in the \xmm\,pipeline).
This cutoff ensures that the \ion{O}{7} features can be measured in most observations ($>$$95\%$).

The contamination due to point sources is removed using the masks generated by the \texttt{cheese} script in the XMM-SAS script \citep{XMMSAS}.
In addition, the potential contamination due to diffuse extragalactic sources is reduced by masking out prominent nearby galaxies and galaxy clusters.
We adopted the Meta-Catalogue of X-ray Detected Clusters of Galaxies \citep[MCXC; ][]{Piffaretti:2011aa} and the \citet{Kourkchi:2017aa} galaxy catalog.
In particular, observations are omitted if more than $50\%$ of the field of view is covered by galaxy clusters or nearby galaxies (see \citetalias{Pan:2024aa} for details).

For each observation, the \ion{O}{7}, \ion{O}{8}, and Fe-L spectral features are extracted in the spectral fitting implemented in XSPEC \citep{XSPEC}.
In total, there are five major components in spectral modeling, i.e., Galactic emissions, local hot bubble (LHB), cosmic X-ray background, soft proton background, and instrumental lines.
The details of these model setups can be found in \citetalias{Pan:2024aa}.
Transitions in the three target spectral features are all set to be zero in the Galactic and LHB emission model (i.e., \texttt{APEC} in XSPEC).
Instead, the three target spectral features are modeled as Gaussian emission lines in the spectral fitting.

\begin{figure*}
    \centering
    \includegraphics[width=0.48\textwidth]{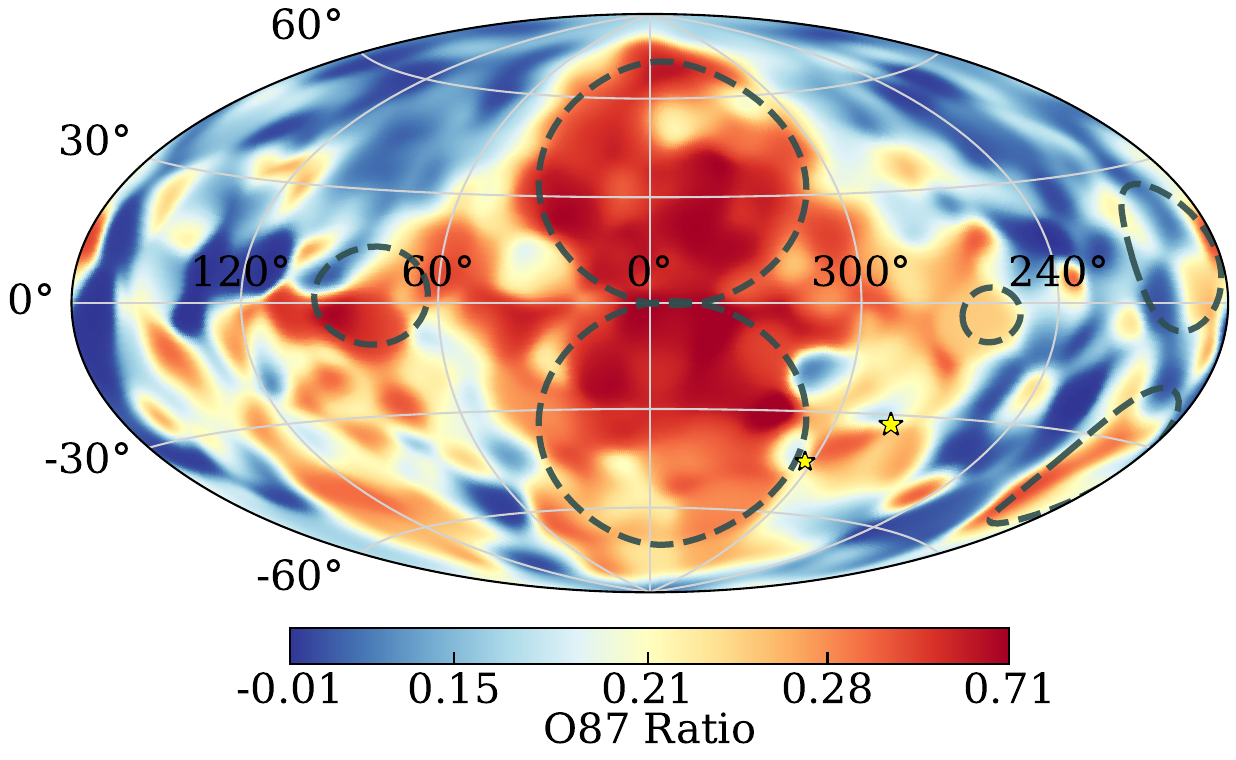}
    \includegraphics[width=0.48\textwidth]{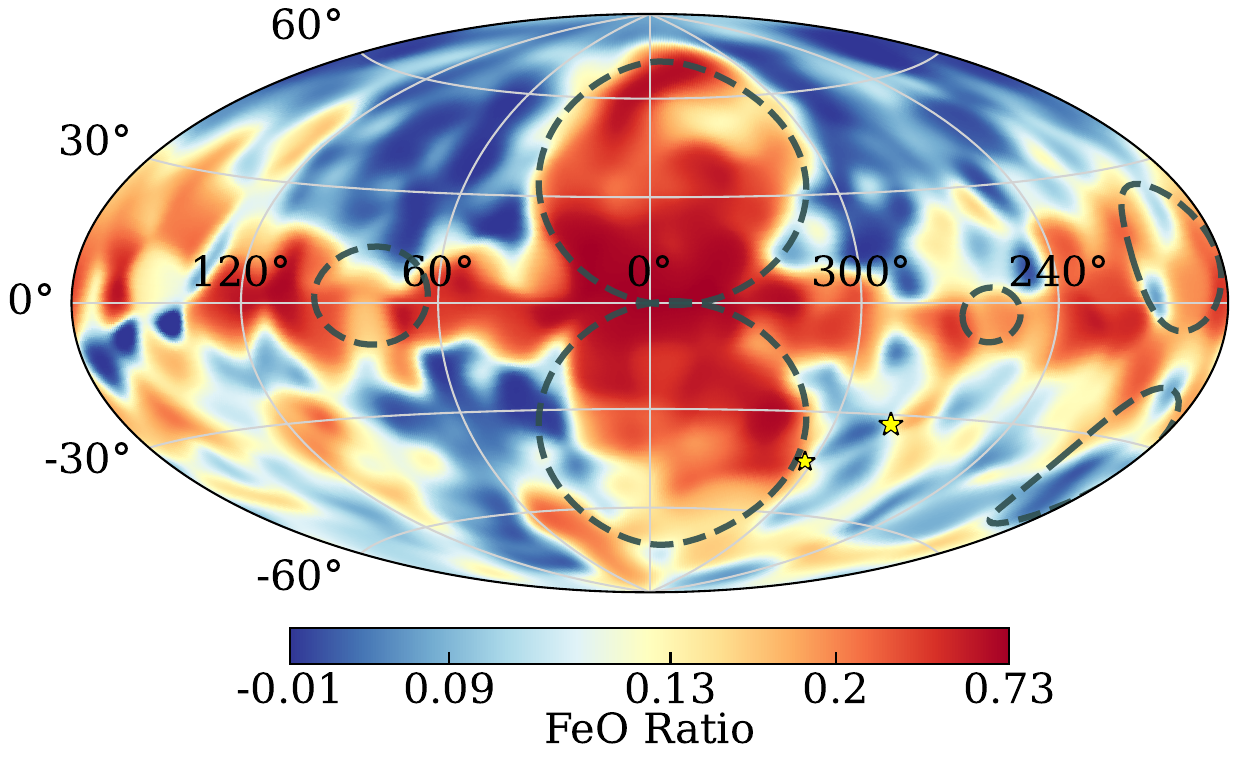}
    \caption{The smoothed maps for line ratios of $I_{\rm OVIII}/I_{\rm OVII}$ (O87; {\it left panel}) and $I_{\rm FeL}/(I_{\rm OVII}+I_{\rm OVIII})$ (FeO; {\it right panel}). For both ratios, higher ratios indicate higher temperatures around the GC. The prominent large-scale structures are marked with dashed lines, including the Fermi and eROSITA bubbles at the GC, Large Magellanic Cloud (LMC; $l, \,b=280^\circ$, $33^\circ$),  Cygnus superbubble ($80^\circ$, $2^\circ$), Vela supernova remnant ($261^\circ$, $-3^\circ$), Monogem supernova remnants ($197^\circ$, $10^\circ$), and Eridanus superbubble ($194^\circ$, $-35^\circ$, see \citetalias{Pan:2024aa} for details). These large-scale structures are not included in the following analyses of the temperature variation.
    }
    \label{fig:2maps}
\end{figure*}

\subsection{The solar wind charge exchange correction}
Although we omitted observations with potentially strong contamination in data reduction, solar wind charge exchange (SWCX) remains a source of all-sky foreground contamination in studies of the Galactic X-ray emission \citep[see a recent review in ][]{Kuntz:2019aa}.
There are two types of SWCX in practice, magnetospheric and heliospheric SWCXs.
The magnetospheric SWCX varies over hours to days \citep{Cravens:2001aa}, while the heliospheric SWCX exhibits significant variation over the solar cycle \citep{Qu:2022ac}.
The contamination due to the magnetospheric SWCX is significantly removed by filtering X-ray flares \citep{Henley:2012aa}.
However, the heliospheric SWCX remains significant in our final sample.
The detailed investigation and modeling of the SWCX dependence on solar (wind) activities and locations (e.g., ecliptic latitude) will be presented in future work.

In this work, we adopt the empirical model described in \citetalias{Pan:2024aa} to correct the heliospheric SWCX.
This model is derived using over 26000 close observational pairs with separations of $\theta\,<\,2^\circ$. At such separations, variations in Galactic emission are considered negligible, supported by the auto-correlation of the emission measure (\citealt{Kaaret:2020aa}) and temperature (Section \ref{sec:acf}).

Using these close pairs, we construct the solar-cycle temporal variation of the SWCX.
This variation is expected to be dominated by the heliospheric SWCX, but will also have contributions due to the magnetospheric SWCX, which varies with the highly-ionized species in the solar winds.

One caveat is that the only relative variation is obtained from the difference between two epochs of each pair without an absolute zero point.
Here, we assume that the SWCX has a zero contribution at the solar minimum to minimize its contribution.
This assumption is supported by detailed SWCX spectral modeling of deep {\it Chandra} and {\it XMM-Newton} observations, suggesting zero contribution during the solar minimum \citep{Huang:2023aa}.

\section{The multi-scale temperature structures of the MW hot CGM}
\label{sec:results}
We extract a sample of three emission features, \ion{O}{7}, \ion{O}{8}, and Fe-L band, after minimizing possible contaminations and correcting the SWCX contribution.
This sample enables a systematic study of the temperature structures in the MW hot gas.
In particular, we define two temperature tracers using the line ratios $I_{\rm OVIII}/I_{\rm OVII}$ (O87 hereafter) and $I_{\rm FeL}/(I_{\rm OVII}+I_{\rm OVIII})$ (FeO) to represent the temperature for the virial phase and the possible extremely hot phase.
In Figure \ref{fig:2maps}, we show two maps of the O87 and FeO ratios, which are generated by smoothing individual observations with a $\sigma = 4^\circ$ Gaussian beam.
This beam size is chosen to smooth the small-scale features and only show global structures of temperature, which will be presented in Section \ref{sec:acf}.

In the temperature range associated with MW hot gas ($\log T/{\rm K} \approx 6-7$), both ratios are positively correlated with the temperature (see \citetalias{Pan:2024aa}).
In this section, we first discuss the temperature implication of the two line-ratio tracers.
Then, we report temperature structures associated with the Fermi and eROSITA bubbles, and examine global temperature variations beyond these bubbles, which are associated with the MW hot gas disk and halo.
In addition, we also extract the auto-correlation and cross-correlation functions of the two temperature variations to investigate the small-scale temperature variation.

\subsection{Comparison between O87 and FeO ratios}
\label{sec:2T_comp}
\begin{figure*}
    \centering
    \includegraphics[width=0.98\textwidth]{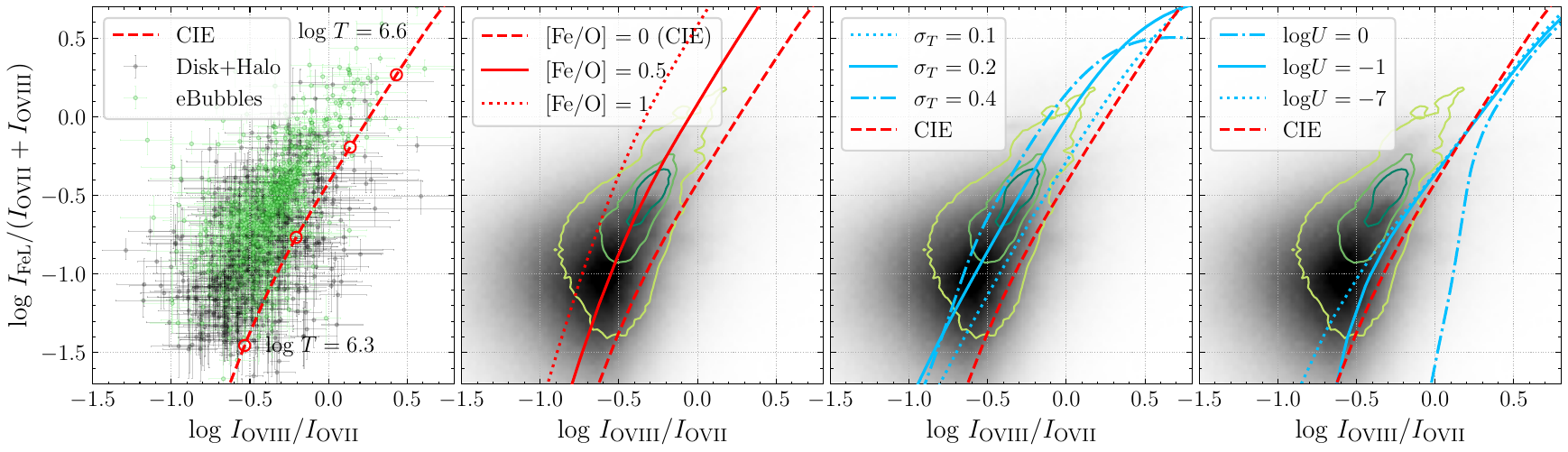}
    \caption{The comparison between two temperature tracers, O87 and FeO ratios.
    Individual measurements of two ratios are shown in the {\it leftmost panel}, which traces the observations with the highest line intensities. 
    The eROSITA bubbles (green dots) exhibit higher ratios compared to the regions beyond the bubbles (black dots).
    The CIE model prediction of the line ratios is shown as the red dashed line and red open circles (i.e., $\log T/{\rm K}=6.3,~ 6.4,~ 6.5$, and 6.6), which are offset from the observed line ratios.
    In the right three panels, we show the contours of the bubbles (green lines) and beyond (grayscale).
    The model predictions can be affected by three factors, the variation of [Fe/O] abundance ({\it middle-left}), the mixing of multi-temperature gas (assumed log-normal distribution with $\sigma_T$; {\it middle-right}), and photoionization ({\it rightmost panel}).
    The multi-temperature scenario is the most likely explanation for the observed line ratios (see Section \ref{sec:2T_comp} for details).
    }
    \label{fig:2T_comp}
\end{figure*}

In Figure \ref{fig:2maps}, although O87 and FeO ratios both show similar large-scale structures, it is of interest to investigate whether these two tracers infer the same temperature of the MW hot gas.
In Figure \ref{fig:2T_comp}, we show the comparison between the two ratios.
In the left panel, we show the pairs with both ratios measured (i.e., \ion{O}{8} and Fe-L detected).
In the following panels, the two-dimensional (2-D) histogram is extracted for all observations including upper limits for these line ratios.
In this 2-D histogram, each observation is represented by 1000 points from the MCMC chain extraction (see \citetalias{Pan:2024aa} for details).
Overall, this distribution is consistent with the distribution of individual measurements, which are both dominated by the observations with detected \ion{O}{8} and Fe-L features.

Next, we compare the observed correlation between the two ratios with the model predictions in Figure \ref{fig:2T_comp}.
In the leftmost panel, it is clear that the majority of measurements are above the CIE model-predicted correlation between O87 and FeO, assuming the solar abundance pattern in \citet{Lodders:2003aa}.
If the hot gas temperature is determined by the O87 ratio, there is an enhancement of Fe-L emission relative to \ion{O}{7} and \ion{O}{8} emissions.
Assuming a solar abundance [Fe/O], the FeO-derived temperature will be higher than the O87-derived temperature.

Such a difference may be explained by three possible scenarios.
First, the [Fe/O] abundance ratio may be super solar.
In the middle-left panel, we show three CIE models with different [Fe/O] abundances.
The red dashed line is the model with the solar [Fe/O], the same as the left panel.
The solid line is the preferred model, where the Fe-L emission is enhanced due to the [Fe/O] abundance of $\approx$$ 0.5$ dex.
However, using the stellar metallicity in the MW as a guide, such a super solar [Fe/O] abundance is unusual in the MW, instead most stars exhibit sub-solar [Fe/O] values $\approx$$ -0.1$ to $-0.2$ dex \citep{Eilers:2022aa}.
In addition, we compare the MW with other galaxies with measured [Fe/O] in X-ray observations \citep{Li:2015aa}.
The late-type galaxies normally have sub-solar [Fe/O] ratios, while super-solar [Fe/O] abundances are typically found in early-type galaxies.
As a spiral galaxy, the MW is expected to have a sub-solar [Fe/O], which is opposite to the needed super-solar $\rm [Fe/O]=0.5$ dex in this scenario.

The second scenario is that the MW hot gas exhibits multiple temperatures instead of a single temperature for each observation.
Considering the FeO temperature is higher than the O87 temperature, the enhanced FeL emission can be due to a super-virial temperature component in the MW hot CGM.
This super-virial component has been proposed to explain the Fe-L enhancement or \ion{Ne}{9} and \ion{Ne}{10} absorption in the QSO field of 1ES 1553 +113 \citep{Das:2019ab, Das:2019aa}.
As shown in Figure \ref{fig:2maps}, this sight line at $l$, $b = 21^\circ$, $43^\circ$ is projected within the northern eROSITA bubble, which has a higher temperature than the typical MW hot gas.
In addition, \citet{Bluem:2022aa} and \citet{Ponti:2023aa} claimed the existence of a super-virial phase for more typical regions of the MW hot CGM using HaloSat data ($|b|>30^\circ$) and eROSITA/eFEDS field data, which are more representative for the hot gas disk or halo.

Here we consider a continuous distribution of temperature instead of a two-temperature model for the MW hot gas.
If the temperature follows a log-normal distribution, the tail extending to the high temperature may explain the enhanced Fe-L emission.
In this scenario, the log-normal distribution is not required, and any distributions with stronger tails at higher temperatures may reproduce the observed relation between O87 and FeO ratios.
Here, we assume the log-normal distribution as an example, and the observed three emission line measurements cannot distinguish this distribution from other possible distributions.
In the middle-right panel of Figure \ref{fig:2T_comp}, we show that the Fe-L emission is more enhanced compared to oxygen emission with a larger scatter of the log-normal distribution ($\sigma_T$).
The best log-normal temperature model has a scatter of $\sigma_T \approx 0.2$ dex, which is also predicted in numerical simulations \citep[e.g.,][]{Vijayan:2022aa}.

In this scenario, there is an upper limit of the enhanced Fe-L emission by increasing $\sigma_T$, which is determined by the width of the emission intensity curve as a function of temperature for these lines (i.e., $\approx 0.4$ dex).
When the $\sigma_T$ is larger than this intrinsic width, the temperature of the gas contributing to the detected emission is completely dominated by the intrinsic width, then increasing $\sigma_T$ will no longer include hotter gas emitting for \ion{O}{7}, \ion{O}{8}, or Fe-L.

In the rightmost panel of Figure \ref{fig:2T_comp}, we also consider the impact of photoionization.
Within 50 kpc from the MW, the ionizing field is dominated by the escaping flux from the MW \citep[e.g.,][]{Fox:2005aa, Bland-Hawthorn:2019aa}. 
However, the MW contribution to the high-energy band is still uncertain in photoionizing \ion{O}{7}, \ion{O}{8}, or Fe-L ions.
Here, we adopt the incident field as HM05 in {\sc cloudy} (i.e., an updated UV background of \citealt{Haardt:2001aa}).
This choice maximizes the photoionization effect on X-ray emitting gas, because the UVB is harder than the MW escaping flux. 
Using the {\sc cloudy} (v17.0; \citealt{Ferland:2017aa}), we predict O87 and FeO ratios as a function of both temperature and ionization parameter ($\log U \equiv \log \Phi_{\rm 912}/cn_{\rm H}$ for ionizing photons).
In this modeling, the normalization of the incident field is less important, while the photoionization solutions are determined by the dimensionless ionization parameter $\log U$.
The $\log U = -7$ model is a reference model when photoionization is negligible, which is expected to be the same as the CIE model.
In practice, these two models are slightly different because {\sc cloudy} uses the \textsc{Chianti} atomic dataset \citep{Del-Zanna:2015aa}, while XSPEC uses the \textsc{AtomDB} \citep{Foster:2012aa}.
Increasing the ionization parameter, the predicted O87 and FeO correlation almost show no variations at $\log U \lesssim -1$.
The $\log U=0$ model significantly reduces the FeO ratio compared to the O87 ratio, because the photoionization is more influential on the relatively lower ionized phase traced by O87.
In observation, the ionization parameter of the X-ray emitting gas is expected to be $ \log U \lesssim -1$ by considering the observed $\log U$ for the UV-absorbing gas \citep[e.g.,][]{Fox:2005aa} and the density difference between UV and X-ray gas \citep[e.g.,][]{Qu:2022ad}.
Therefore, we conclude photoionization is unlikely to explain the observed enhancement of Fe-L emission.

Current low-spectral-resolution observations cannot provide strong constraints on these scenarios, especially the first two possibilities.
Instead, better constraints on the temperature distribution and abundance variation are expected to be obtained by future missions such as Athena, HUBS, and LEM \citep{Athena, HUBS, LEM, Bregman:2023aa}.

\begin{figure*}
    \centering
    \includegraphics[width=0.90\textwidth]{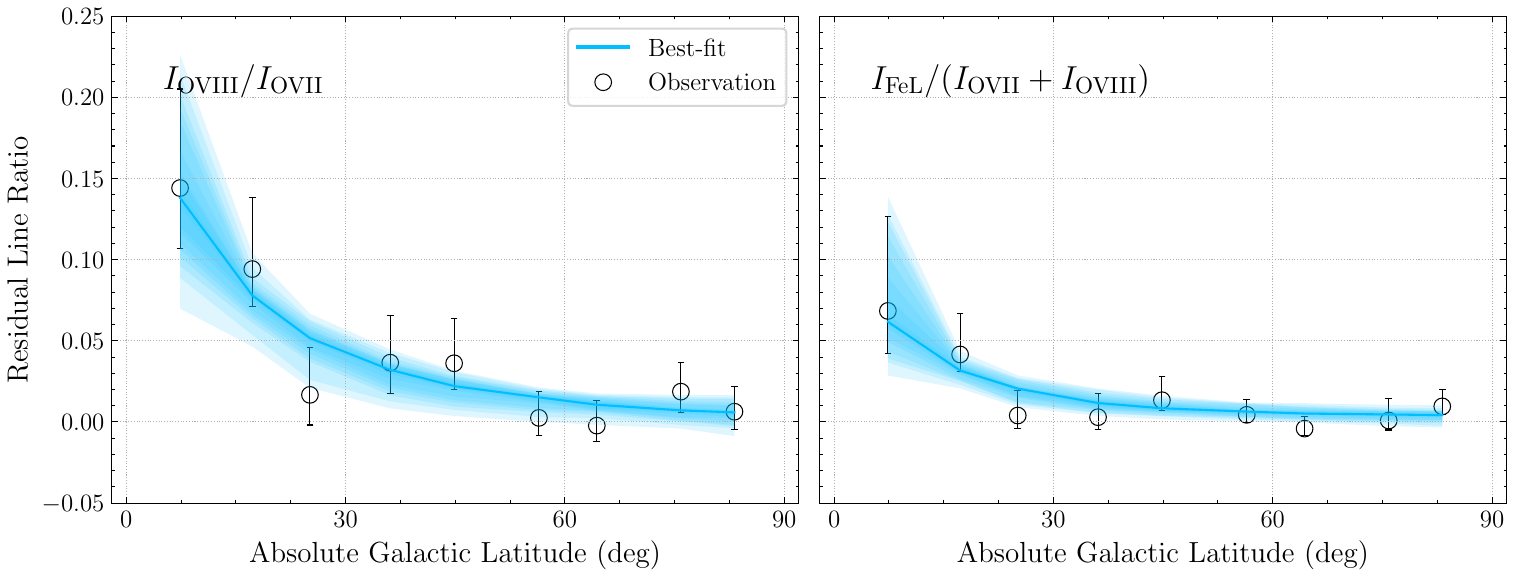}
    \caption{The O87 and FeO dependence on the absolute Galactic latitude in 10$^\circ$ bins, showing the temperature variation of the hot gas disk. 
    After removing the halo component using the dependence on $\tgc$, the residual line ratios show the disk component of the temperature variation. The extension along the $z$-direction is modeled using an exponential density profile (blue lines), which leads to the best-fit scale angles of ${20^\circ}_{-8^\circ}^{+31^\circ}$ and ${13^\circ}_{-6^\circ}^{+19^\circ}$ for O87 and FeO, respectively.
    }
    \label{fig:disk}
\end{figure*}

\subsection{The eROSITA bubbles}
\label{sec:bubbles}

In Figure \ref{fig:2maps}, we show the two smoothed maps, where auxiliary lines are also plotted to indicate the position of the Fermi and eROSITA bubbles.
In both maps, the eROSITA bubbles show higher temperatures than the all-sky average, while the Fermi bubbles are not significantly different from the other regions within the eROSITA bubbles. 
Within the Fermi bubbles, the gas is relativistic, which does not emit in the X-ray band \citep[e.g.,][]{Yang:2022aa}.
Therefore, in the following analyses, we only consider the hot gas temperature in the eROSITA bubbles.

In Figure \ref{fig:2T_comp}, we also compare the O87 and FeO ratios in the eROSITA bubbles.
Overall, the bubbles exhibit higher O87 and FeO ratios with median values of $\approx$$ 0.3$ and $0.3$ (i.e., $-0.5$ dex).
Assuming the solar [Fe/O] abundance, these two ratios can be explained by a log-normal temperature distribution of $\log T/{\rm K} = 6.4$ and $\sigma_T=0.2$ dex.
This temperature distribution is consistent with previous studies suggesting a hotter phase is needed around the Bubbles \citep[e.g.,][]{Miller:2016aa, Das:2019ab, Das:2019aa}.

Both O87 and FeO maps indicate higher temperatures within the eROSITA bubbles, but there are still differences between the two tracers.
The O87 ratios in the bubbles are not significantly different from nearby positions, especially at lower Galactic latitudes.
The O87 map seems to be more spherical, where the temperature is only a function of the angular separation to the GC, while the FeO map shows a clearer bubble shape associated with the eROSITA bubbles.
In addition, the FeO ratio shows a more significant disk component at low latitudes.

\subsection{The hot gas disk}
\label{sec:disk}
Previous studies investigated the spatial distribution of the hot gas using the dependence of X-ray emission intensity on galactic longitudes and latitudes.
In general, the MW hot gas can be approximated by a gaseous disk and a halo.
Existing studies found a scale height of $1-2$ kpc in an exponential disk model \citep[e.g.,][]{Li:2017aa, Nakashima:2018aa, Kaaret:2020aa}.
In different studies, the halo component can be modeled under hydrostatic equilibrium assumption or simply a cored power law \citep[e.g.,][]{Faerman:2017aa, Li:2017aa}.
Using the new sample introduced in \citetalias{Pan:2024aa}, the density distribution will be further investigated in future work.

Here, we consider whether the temperature of the MW hot gas can be decomposed into disk and halo components.
In the following analysis, the eROSITA bubbles are excluded using the region shown in Figure \ref{fig:2maps}.
An empirical test of the potential disk component is investigating the X-ray emission dependence on the Galactic latitude.
Assuming a spherical distribution of the halo, the temperature (i.e., O87 and FeO ratios) should only depend on the distance to the GC ($\tgc$).
However, in a disk-like model, the temperatures measured at the same $\tgc$ also have a dependence on the Galactic latitude.
To exclude the contribution due to the spherical CGM component, we subtract a one-dimensional median radial profile as a function of $\tgc$, which is fitted as an exponential function.
Then, we select observations with Galactic longitudes at $80^\circ< l < 100^\circ$ to minimize the impact of the radial variation along the midplane of the disk component.
The disk component's dependence on Galactic latitude varies across different longitudes, which has been seen in UV absorption line studies for the warm gas at $\log T/{\rm K}\approx 4-5$ \citep[e.g.,][]{Qu:2019ab}. 
In Figure \ref{fig:disk}, we show the residual ratio dependence on the Galactic latitude at $80^\circ< l < 100^\circ$, which shows significant declines to higher latitudes for both O87 and FeO ratios.
Such a variation indicates that the hot gas has a higher temperature around the midplane of the disk.

Furthermore, we use the exponential function of $A \exp (-2x/\phi_0)+C$ to fit the vertical variation of both ratios, where $\phi_0$ is the scale angle for the extension in the $z$-direction.
This format is adopted because the exponential disk is normally assumed for the warm-hot gas density in the MW disk \citep[e.g.,][]{Savage:2009aa, Li:2017aa}, and the factor of 2 is because of the squared dependence of emission on the density.
In this function, $C$ is the free parameter for the background level, because we subtract the radial profile as a function of distance to the GC, which does not guarantee a zero background level.
The fitting is performed in the Bayesian framework using the \texttt{emcee} implementation (\citealt{Foreman-Mackey:2013aa}), which has a likelihood function of
\begin{eqnarray}
    p(A, r_0, C) = &\Pi_i\int\frac{1}{\sigma_{y, i}} \times {\rm exp} \left(- \frac{[y_{{\rm m}, i}-y_i]^2}{2\sigma^2_{y, i}}\right) {\rm d} t,
\end{eqnarray}
where $y_{{\rm m}, i}(A, \phi_0, C)$ is the model prediction, while $y_i$ and $\sigma_{y,i}$ are the residual temperature ratio and uncertainty in each $10^\circ$ bin.
The best-fit parameters are $A = 0.20_{-0.05}^{+0.11}$, $\phi_0 = {20^\circ}_{-8^\circ}^{+31^\circ}$, and $C=0.00_{-0.03}^{+0.01}$ for the O87 ratio, while $A = 0.11_{-0.05}^{+0.19}$, $\phi_0 = {13^\circ}_{-6^\circ}^{+19^\circ}$, and $C=0.00_{-0.01}^{+0.01}$ for the FeO ratio.
These two scale angles are consistent with each other within the $1\sigma$ uncertainty, while the FeO ratio may be more extended than O87 in the $z$-direction, but only using the data at $80^\circ<l<100^\circ$ cannot solidly distinguish the difference.

To constrain the disk component in temperature using all observations, a more systematic model including both disk radial and vertical variation should be considered for both temperature and density spatial variations.
Such a sophisticated model requires a better understanding of the SWCX especially the temporal variation in hours to days and spatial variation of the SWCX, which is discussed in \citetalias{Pan:2024aa}, and will be more investigated in future works.

\begin{figure*}
    \centering
    \includegraphics[width=0.98\textwidth]{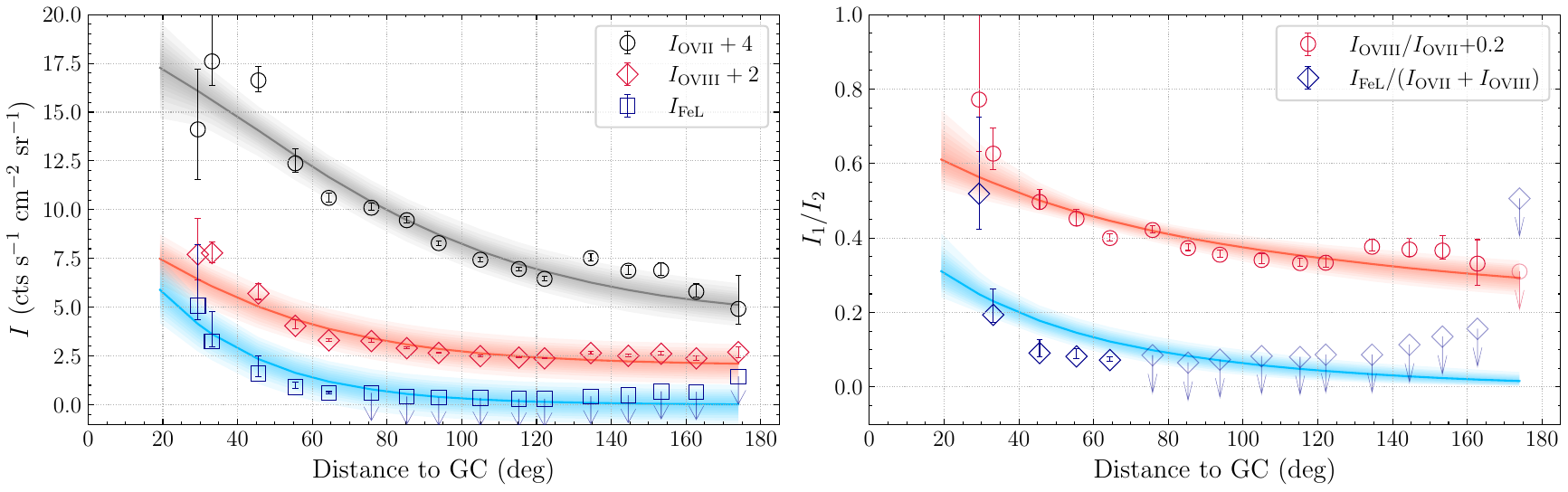}
    \caption{The line intensity ({\it left panel}) and line ratio ({\it right panel}) dependence on the distance to GC ($\tgc$) in 10$^\circ$ bins.
    In this plot, the eROSITA bubbles are removed to avoid contamination.
    The lines are the best joint-fit models for line intensities, shown in Figure \ref{fig:radial_T}.
    }
    \label{fig:radial_ratio}
\end{figure*}

\subsection{The hot gas halo}
\label{sec:halo}
Besides the disk component, we also investigate the temperature variation associated with the halo component.
In Figure \ref{fig:radial_ratio}, we show the radial profile of \ion{O}{7}, \ion{O}{8}, and Fe-L emissions as functions of $\tgc$.
From the GC to the anti-GC, the spectral features \ion{O}{7}, \ion{O}{8}, and Fe-L show declines of $\approx$$ 11$ L.U., 5 L.U., and 5 L.U. for \ion{O}{7}, \ion{O}{8}, and Fe-L, respectively.

\begin{figure*}
    \centering
    \includegraphics[width=0.98\textwidth]{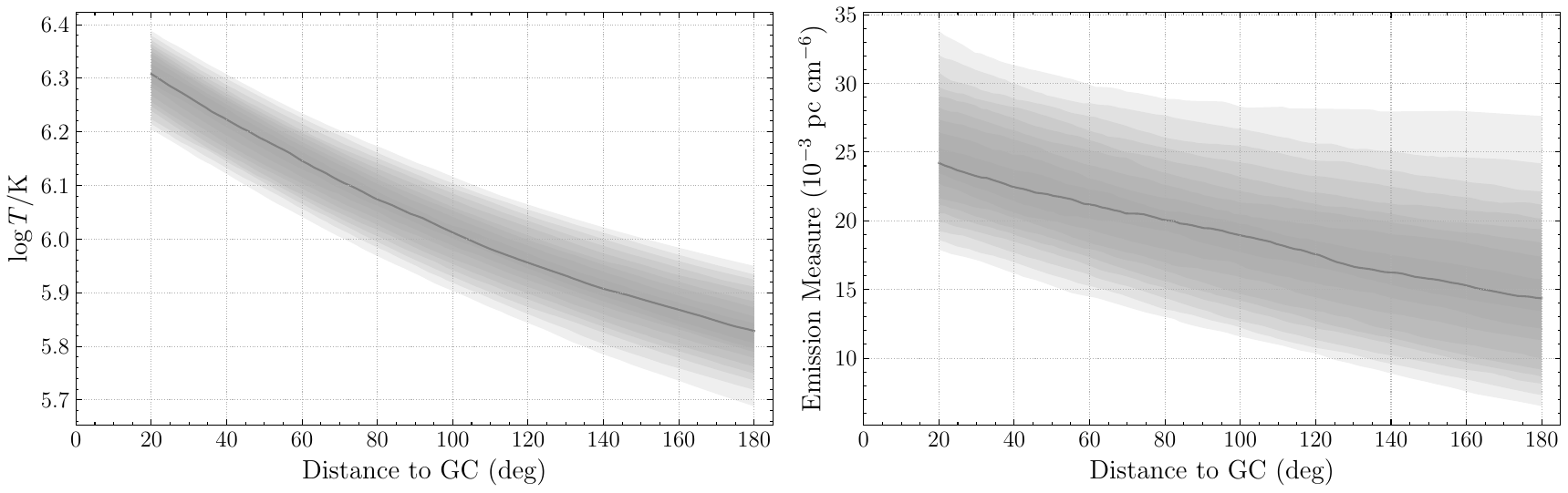}
    \caption{The best-fit temperature ({\it left panel}) and emission measure ({\it right panel}) decline from the GC to the anti-GC, showing the spatial variation of the hot gas halo. Dominated by the decline of line ratios shown in Figure \ref{fig:radial_ratio}, the temperature decreases significantly from $\log\,T/{\rm K}\approx6.3$ to $5.8$.
    The emission measure is consistent with zero at $\tgc > 160^\circ$, which is mainly affected by the spatial variation of the SWCX (see Section \ref{sec:halo} and \citet{Pan:2024aa} for details).}
    \label{fig:radial_T}
\end{figure*}

In addition, the \ion{O}{7} emission seems to exhibit a flat or lower emission within $45^\circ$, which may be due to the increasing temperature.
Therefore, we further show the O87 and FeO ratio dependences on $\tgc$ tracing the temperature in Figure \ref{fig:radial_T}.
The O87 ratio decreases from $0.4$ to $0.1$, while FeO decreases from $0.4$ to zero, indicating the temperature decline from the GC to the anti-GC.
The decline of the temperature may also reduce the intensities of the three emission features, together with the decrease of the emission measure ($E\equiv n^2 L$).

Next, we decompose the declines of the temperature and the emission measure by fitting the decline of the \ion{O}{7}, \ion{O}{8}, and Fe-L emission simultaneously.
We consider an empirical model with the temperature and emission measure dependences on $\tgc$.
The dependences are assumed to be exponential as $A\exp(-x/r_0)+C$ for both the temperature ($\log T/{\rm K}$) and emission measure ($E$).
We further assume a log-normal distribution for the temperature, with the $\sigma_T$ as a free parameter constant for all directions.
In addition, we also include an artificial patchiness parameter ($\sigma_p$) to account for unresolved uncertainties in the measurements (e.g., the SWCX correction).
Then, we use this model to simultaneously reproduce the observed three spectral emission features, where the line emissivities are adopted from {\sc AtomDB} \citep{Foster:2012aa}.
The best-fit temperature and emission measure ($E$) profiles are
\begin{equation}
    \log T(\tgc)/{\rm K} = (0.9\pm 0.1) \exp(-\frac{\tgc}{180^{\circ}}) + 5.4\pm0.2,
\end{equation}
and 
\begin{equation}
    10^3 E(\tgc) = (21\pm 13) \exp(-\frac{\tgc}{180^{\circ}}) - 12\pm 11 {\rm~ pc~cm^{-6}},
\end{equation}
where $r_0$ is fixed at $180^\circ$ for both profiles because it is highly degenerate with $A$ and $C$, and varying it as a free parameter leads to $2\sigma$ lower limit of  $80^\circ$.
The best-fit $\sigma_T$ is $0.23\pm0.04$, which is consistent with the $\sigma_T \approx 0.2$ determined by the observations with the strongest emission (Figure \ref{fig:2T_comp}).
The patchiness parameter $\sigma_{\rm p}$ is $0.6\pm0.1$, which is expected to be the residual in the SWCX correction and light-of-sight variations.

The model is also plotted in Figure \ref{fig:radial_ratio} to compare with the observation data.
Because we fit the absolute emission in this empirical model, the comparison with the O87 and FeO ratios is a self-consistent check for the fit of the temperature.
The O87 ratios are reproduced well at all $\tgc$, rather than the FeO ratios within $\approx$$40^\circ$, which suggests that \ion{O}{7} and \ion{O}{8} emissions dominate the fitting because of their high S/N values.

Figure \ref{fig:radial_T} shows the model-constrained temperature and emission measure profiles.
From the GC to anti-GC, the temperature decreases from $\log T/{\rm K}\approx 6.3$ (i.e., 0.18 keV) to $5.8$ (0.07 keV), and the emission measure also shows a decline from $24\times 10^{-3}$ to $14\times 10^{-3} \rm ~pc ~cm^{-6}$.
At $\tgc>150^\circ$, the emission measure exhibits negative values, which is consistent with zero, due to the uncertainties associated with SWCX correction.
The comparison with previous studies will be discussed in Section \ref{sec:dis_temperature}.

\begin{figure*}
    \centering
    \includegraphics[width=0.98\textwidth]{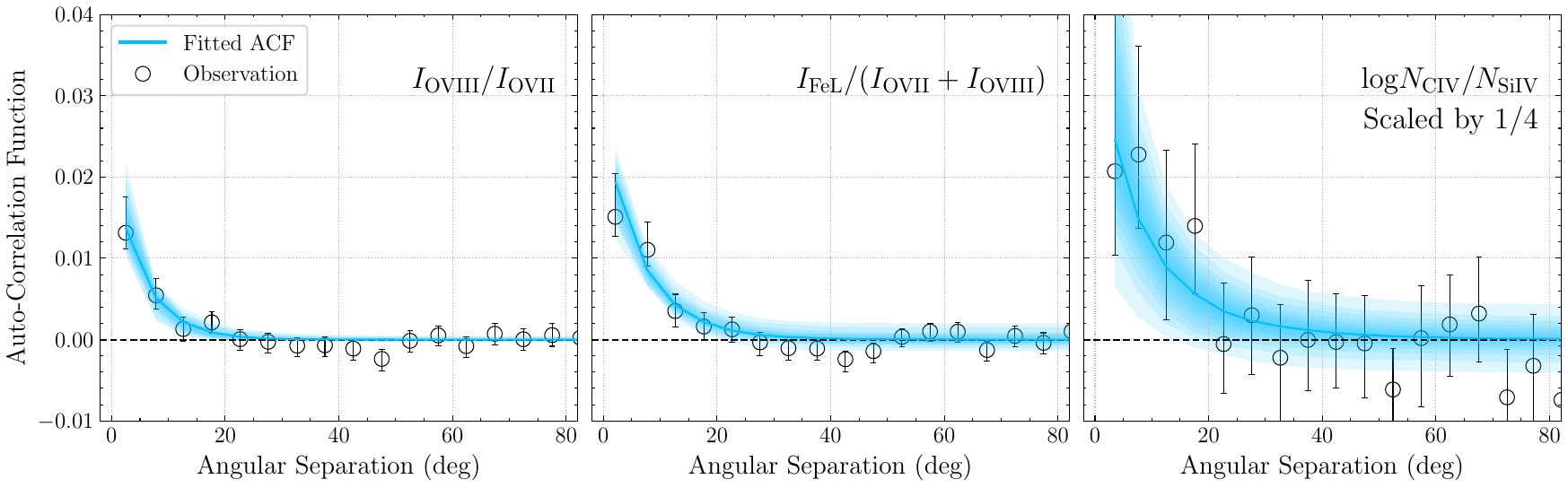}
    \caption{The auto-correlation function of the O87 ({\it left panel}), the FeO ({\it middle panel}), and the C4Si4 ratios (intermediate phase; {\it right panel}).
    Fitted with exponential functions, correlation angles are $5.2^\circ\pm 1.4^\circ$, $7.0^\circ\pm 1.7^\circ$, and ${10.2^\circ}_{-5.1^\circ}^{+7.6^\circ}$, respectively.
    The \ion{C}{4} and \ion{Si}{4} data is adopted from \citet{Qu:2022aa}.}
    \label{fig:acf}
\end{figure*}

The decomposition of the declines of temperature and emission measure reveals the complexity of the large-scale structure in the MW hot gas.
A three-dimensional decomposition of the temperature and density distribution will be done after the SWCX is better modeled.
 
\subsection{The small-scale variations}
\label{sec:acf}
We further investigate the small-scale structures in the temperature variation of the hot gas in the MW.
After excluding the eROSITA bubbles, we calculate the auto-correlation function (ACF) for the disk and halo emission.
In the ACF, there is a large-scale correction beyond $\gtrsim$$ 100^\circ$, which is due to the global variation of the disk and halo.
In addition, there are also small-scale correlations, due to the local variation of the hot gas temperature, shown in Figure \ref{fig:acf} for both O87 and FeO ratios.
The ACF is normalized by the variance, and we did not include pairs of any individual observations with themselves. 
The maximum positive correlation occurs at the angular separation ($\theta$) approaching $0^\circ$, which is determined by the relative contribution of measurement uncertainty to the total observed scatter, so it is not the unity at $\theta=0^\circ$.
It is clear that between $\theta=20^\circ-80^\circ$, there is no significant correlation, while within $20^\circ$, strong positive correlations are found for both ratios.

This positive correlation indicates the typical angular size of the temperature variation.
Then, we use the exponential function $A\exp(-\theta / \theta_0)$ to obtain the correlation scale, where $A$ and $\theta_0$ are free parameters.
The best-fit $\theta_0$ are $5.2^\circ \pm 1.4^\circ$ and $7.0^\circ\pm1.7^\circ$ for O87 and FeO, respectively.
These two correlation angular sizes are consistent with each other within $1 \sigma$.

Next, we calculate the cross-correlation function (CCF) between O87 and FeO ratios (Figure \ref{fig:ccf}).
The positive correlation is shown at a small scale ($<$$ 10^\circ$), which is consistent with the auto-correction function.
This positive cross-correlation suggests that the multi-temperature hot gas may share the same origins.

\section{Implication and Discussion}
\label{sec:dis}
In Section \ref{sec:results}, we examine the multi-scale temperature structures in the MW hot gas.
First, we show that the hot gas temperature has three large-scale features: the bubbles, disk, and halo, which are similar to the density distribution.
Furthermore, we report the enhancement of the temperature around the GC, which is contributed by both the eROSITA bubbles and more extended disk or halo components.
In addition, the hot gas shows variations with an angular size of $\lesssim$$ 10^\circ$ at a small scale.
In this section, we discuss the implications of these findings.

\subsection{Temperature structures of the MW hot gas beyond the eROSITA bubbles}
\label{sec:dis_temperature}

In Section \ref{sec:results}, we found the large-scale hot gas exhibits disk and halo components similar to the density distribution.
The disk component is suggested because of higher values for both O87 and FeO ratios at low Galactic latitudes (Figure \ref{fig:disk}).
We further found that the scale angle is about $\approx$$ 20^\circ$, which can be converted into physical scale height when the distance is known.
The angular size of $20^\circ$ is 1.7 kpc assuming an average distance of 5 kpc for the X-ray emission due to hot gas, which is estimated from the density distribution obtained from previous studies \citep[e.g.,][]{Miller:2015aa}.
This estimated scale height is consistent with scale heights of the density distribution of $\approx$$ 1.0-2.0$ kpc \citep[e.g.,][]{Li:2017aa, Nakashima:2018aa, Kaaret:2020aa}.
These scale heights of the X-ray emitting gas for both density and temperature are lower than the UV ions of $\approx$$ 3.0$ kpc, assuming the exponential density distribution along the $z$-direction \citep[e.g.,][]{Savage:2009aa, Qu:2019ab}.

\begin{figure*}
    \centering
    \includegraphics[width=0.98\textwidth]{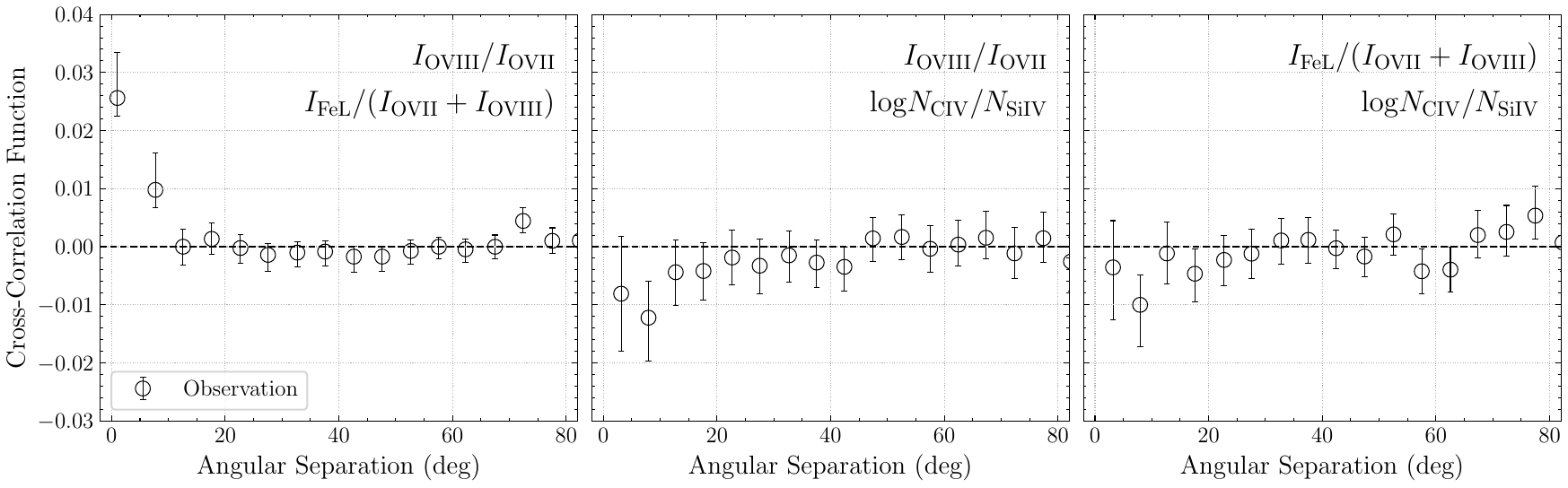}
    \caption{The cross-correlation functions between the O87, FeO, and C4Si4 ratios.
    The correlated ratios are marked in the top right corner of each panel.
    The two ratios in the X-ray band are positively correlated at the small separation of $\lesssim 10^\circ$, suggesting physical connections.
    The C4Si4 ratio tracing the intermediate ionization phase is negatively correlated with the hot gas at the small separation, showing a total significance of $3.0\sigma$ with both O87 and FeO.
    }
    \label{fig:ccf}
\end{figure*}

The temperature of the halo component is investigated based on its dependence on the angular distance to the GC, after removing the contribution of the eROSITA bubbles.
Both O87 and FeO ratios exhibit significant declines, which is explained by the temperature decline.
By fitting the intensity of \ion{O}{7}, \ion{O}{8}, and Fe-L emissions, we decompose the declines of the temperature and emission measure.
This temperature decline is different from previous studies reporting an almost constant temperature of $\log\,T/{\rm K}\approx6.2$ \citep[0.2 keV;][]{Henley:2013aa, Kaaret:2020aa}.
Recently, \citet{Bluem:2022aa} decomposes the MW hot gas emission at $|b|>30^\circ$ into two phases with temperatures of $\log\,T/{\rm K}\approx6.2$ and $6.7-7.0$, and both phases show dependence on $\tgc$.
In addition, we also get the best-fit scatter of $0.23\pm 0.04$ in the log-normal temperature.
The region at $\tgc > 60^\circ$ dominates this scatter, consistent with the temperature scatter in the eROSITA bubbles ($\approx$$ 0.2$ dex).

At last, we examine the small-scale structure by analyzing the auto-correlation of O87 and  FeO ratios, and the cross-correlation between these two ratios. 
Assuming a distance of 5 kpc, the correlation angle of $5^\circ$ can be converted into a physical size of $\approx$$ 400$ pc.
The X-ray bubbles associated with the star-forming region or supernova remnant are about $50-100$ pc \citep[e.g.][]{Lopez:2014aa, Asvarov:2014aa}.
The larger physical size of the temperature structure suggests that the detected temperature variation may be due to the evolved star-forming regions.
This is consistent with the scenario that X-ray emission is preferentially dominated by star-forming regions rather than extended diffuse thick disk suggested by \citet{Kaaret:2020aa}, which found the hot gas emission is better modeled following the distribution of the molecular gas than a normally-assumed exponential disk.

\subsection{Comparisons with ionized species traced by UV lines}
\label{sec:uv}
We further investigate the connections between the X-ray emitting gas and the warm gas traced by intermediate ions, such as the \ion{Si}{4}, \ion{C}{4}, and \ion{O}{6} \citep[e.g.,][]{Savage:2003aa, Sembach:2003aa, Lehner:2010aa, Zheng:2019aa, Qu:2019ab}.
The warm gas with lower ionization than the X-ray-emitting gas is expected to be a transient gaseous phase surrounding the MW disk \citep{Savage:2009aa} or associated with the eROSITA bubbles \citep{Soto:2023aa}.
Currently, the ionization mechanism of the warm gas is still unclear. 
It can be either photoionized with an ionization parameter of $\log U \approx -2$ or collisionally ionized at a temperature of $\log T/{\rm K}\approx 5$.
Here, we use $\log N_{\rm CIV}/N_{\rm SiIV}$ as a tracer of the variation of the warm gas.
The \ion{C}{4} and \ion{Si}{4} data is adopted from \citet{Qu:2022aa}.
To produce \ion{C}{4}, it requires ionizing photons of energy exceeding $47.8$ eV. In contrast, it takes $45.1$ eV to destroy \ion{Si}{4}.
Therefore,  $N_{\rm CIV}/ N_{\rm SiIV}$ ratio (hereafter C4Si4) provides a measure of the intermediate ionization state of this warm phase.
A higher C4Si4 ratio would indicate a higher ionization state of the intermediate phase.

In the right panels of Figures \ref{fig:acf} and \ref{fig:ccf}, we show the ACF of the C4Si4 ratio, and its CCFs with the O87 and FeO ratios, respectively.
The C4Si4 ratio has an auto-correlation angular scale of ${10.2^\circ}_{-5.1^\circ}^{+7.6^\circ}$, which is slightly larger than the O87 and FeO ratios, but still within the $\approx$$ 1 \sigma$ uncertainties.
At 5 kpc, this angle corresponds to a physical size of $0.8_{-0.4}^{+0.6}$ kpc.
The CCFs with the O87 and FeO ratios exhibit a negative correlation within $20^\circ$ with a significance of $\approx$$ 2.5 \sigma$ and $1.6 \sigma$, respectively.
Totally, the negative correlation exhibits a significance of 3.0$\sigma$ between the strength of the UV-absorbing gas and X-ray emitting gas.

The negative correlation between C4Si4 and O87 suggests that the warm gas probed in UV absorption is associated with the X-ray emitting gas.
However, caveats exist in explaining this association.
First, abundance variations could modify the cross-correlation between the X-ray and UV ion ratios.
As an $\alpha$-element, silicon is produced in massive stars, while carbon also has a considerable contribution from the winds of AGB stars.
In star-forming regions or supernova remnants, [C/Si] may be different from other regions, traced by stars and affected by dust depletion \citep{Eilers:2022aa, Savage:1996aa}.
Second, ionization mechanisms of \ion{C}{4} and \ion{Si}{4} are still uncertain \citep[e.g.,][]{Sembach:2003aa, Fox:2004aa, Collins:2004aa}, making it hard to examine more physical connection between the warm gas and hot gas.

\section{Summary}
\label{sec:summary}
In this work, we report the multi-scale temperature structures in the MW hot gas, using a newly composed sample of 5418 observations reported in \citetalias{Pan:2024aa}.
In particular, we define two line ratios, O87
and FeO, tracing the temperature of X-ray-emitting gas. 
The major findings are summarized below.
\begin{itemize}
    \item There are three major components of the temperature variation, the eROSITA bubbles, the hot gas disk, and the hot halo, which is similar to the density distribution (Figure \ref{fig:2maps}).
    \item The two temperature tracers (i.e., O87 and FeO) cannot be explained by single-temperature CIE models (Figure \ref{fig:2T_comp}). We suggest that the temperature scatter in the MW hot gas can explain the observed relation between O87 and FeO.
    \item In the eROSITA bubbles, the O87 and FeO ratios are both $\approx$$ 0.3$, which can be explained by a log-normal temperature distribution of $\log T/{\rm K} \approx 6.4$ with a scatter of $\sigma_{T}\approx 0.2$ dex.
    This temperature scatter in the Bubbles is consistent with the values in the rest regions of $0.23\pm 0.04$ dex.
    \item The disk component is shown by the ratio dependence on Galactic latitudes, showing the higher temperature at low latitudes (Figure \ref{fig:disk}). Assuming a distance of 5 kpc, the scale height is estimated to be $\approx$$ 2$ kpc, similar to the scale height of hot gas density distribution.
    \item Both O87 and FeO ratios decline from the GC to the anti-GC, showing a variation in the temperature of the halo component  (Figure \ref{fig:radial_ratio}). We further decompose the declines of temperature from $\log T/{\rm K}\approx 6.3$ to $5.8$ and emission measure from $\approx$$ 24\times 10^{-3}\rm ~pc~cm^{-3}$ to $14\times 10^{-3}\rm ~pc~cm^{-3}$ (Figure \ref{fig:radial_T}).
    \item The small-scale structure is investigated by measuring the ACFs of both O87 and FeO ratios, which show correlations within $\approx$$ 5^\circ$ (Figure \ref{fig:acf}). This auto-correlation is consistent with the intermediate phase gas traced by \ion{Si}{4} and \ion{C}{4}. The positive CCF between O87 and FeO ratios indicates different phases in the hot gas are physically correlated with each other (Figure \ref{fig:ccf}).\\
    \item We also analyze the correlation between the X-ray temperature with the ionization state of warm gas characterized by $\log N_{\rm CIV}/N_{\rm SiIV}$. The negative correlation between the X-ray temperature and the intermediate ionization state suggests that the warm gas is associated with the X-ray emitting gas, while the physical reason is still unclear  (Section \ref{sec:uv} and Figure \ref{fig:ccf}).
\end{itemize}

~\\
We thank the anonymous referee for the valuable comments and suggestions that significantly improved our work. 
The authors thank Yakov Faerman, Jiangtao Li, and Hsiao-Wen Chen for their thoughtful suggestions.
JFL acknowledges support from the NSFC through grant Nos. 11988101 and 11933004, and support from the New Cornerstone Science Foundation through the New Cornerstone Investigator Program and the XPLORER PRIZE.
JNB acknowledges support from the University of Michigan.
This research is based on observations obtained with \xmm, an ESA science mission with instruments and contributions directly funded by ESA Member States and NASA.

\bibliography{sample631}{}
\bibliographystyle{aasjournal}



\end{document}